\documentstyle[twoside,fleqn,espcrc2,epsf]{article}

\newcommand{\beq}{\begin{equation}}
\newcommand{\eeq}{\end{equation}}
\newcommand{\bea}{\begin{eqnarray}}
\newcommand{\eea}{\end{eqnarray}}
\newcommand{\ol}[1]{\overline{#1}}

\newcommand{\qbq}{\langle\ol{q}q\rangle}
\newcommand{\mres}{m_{\rm res}}

\newcommand{\dmqbq}{\delta m_{\langle\ol{q}q\rangle}}
\newcommand{\corrpp}{\langle \pi^a(x) \pi^a(0) \rangle}
\newcommand{\corrss}{\langle \sigma(x) \sigma(0) \rangle_{\rm c}}
\newcommand{\corraa}{\langle A_0^a(x) A_0^a(0) \rangle}

\title{ 
Light hadronic physics using domain wall fermions
in quenched \\ lattice 
QCD\thanks{Work done in
collaboration with T.~Blum, P.~Chen, N.~Christ, C.~Cristian,
C.~Dawson, G.~Fleming, A.~Kaehler, X.~Liao, G.~Liu,
C.~Malureanu, R.~Mawhinney, S.~Ohta, G.~Siegert,
A.~Soni, C.~Sui, P.~Vranas, L.~Wu, and Y.~Zhestkov 
(RIKEN/BNL/CU Collaboration)
} % end thanks
} % end title

\author{
Matthew Wingate\address{RIKEN BNL Research Center,
	Brookhaven National Laboratory, Upton, NY 11973, USA}
}

\begin{document}

\begin{abstract}
In the past year domain wall fermion simulations have 
moved from exploratory stages to the point where
systematic effects can be studied with different gauge 
couplings, volumes, and lengths in the fifth dimension.
Results are presented here for the chiral condensate,
the light hadron spectrum, and the strange quark mass.
We focus especially on the pseudoscalar meson mass
and show that, in small volume, the correlators used to compute it can
be contaminated to different degrees by topological zero modes.
In large volume a nonlinear
extrapolation to the chiral limit, e.g.\ as expected from
quenched chiral perturbation theory, is needed in order
to have a consistent picture of low energy chiral symmetry
breaking effects.
\end{abstract}

\maketitle

%%%%%
\section{INTRODUCTION}

The RBC Collaboration has recently reported results based on
quenched QCD simulations with domain wall fermions on
lattices of various volumes and spacings~\cite{Blum:2000kn}.
This talk summarizes our understanding of the 
behavior of the pseudoscalar meson mass as a function of 
input quark mass, scaling of the nucleon--rho mass ratio,
and strange quark mass.  Another set of quenched
simulation results with domain wall fermions have been reported
by the CP-PACS Collaboration~\cite{AliKhan:2000iv}.

%%%%%
\section{CHIRAL CONDENSATE}

A Dirac operator $D$ which has a well-defined index
should lead to a chiral condensate which satisfies
the following Banks--Casher relation (for finite volume) 
\beq
-\qbq =
\frac{1}{12V} \frac { \langle |\nu|\rangle }{m} +
\frac{m}{12V} \left \langle \sum_{i, \lambda_i \ne 0}
          \frac{1}{\lambda_i^2 + m^2} \right \rangle
\label{eq:BC}
\eeq
where $\lambda_i$ are eigenvalues of the massless Dirac operator
and $\nu$ is the index of $D$.
Note that these zero modes are chiral and
correspond to units of topological charge in the continuum.
In full QCD simulations configurations which 
support eigenvalues equal to zero are suppressed during Monte
Carlo updating due to the fermion determinant in the
Boltzmann weight.  This suppression is absent
in the quenched approximation.  Consequently, a prominent
$1/m$ divergence should appear in $\qbq$ for quenched 
simulations in finite volumes.  Only when one uses a
fermion discretization which admits a non-zero index can one observe
this divergence, so the first signal was
seen with domain wall fermions
\cite{Fleming:1999cc,Kaehler:1999kg}.

\begin{figure}[t]
%\epsfxsize=\hsize
%\epsfbox{qbq_b57k165ns32.eps}
\vspace{4.0cm}
\includegraphics{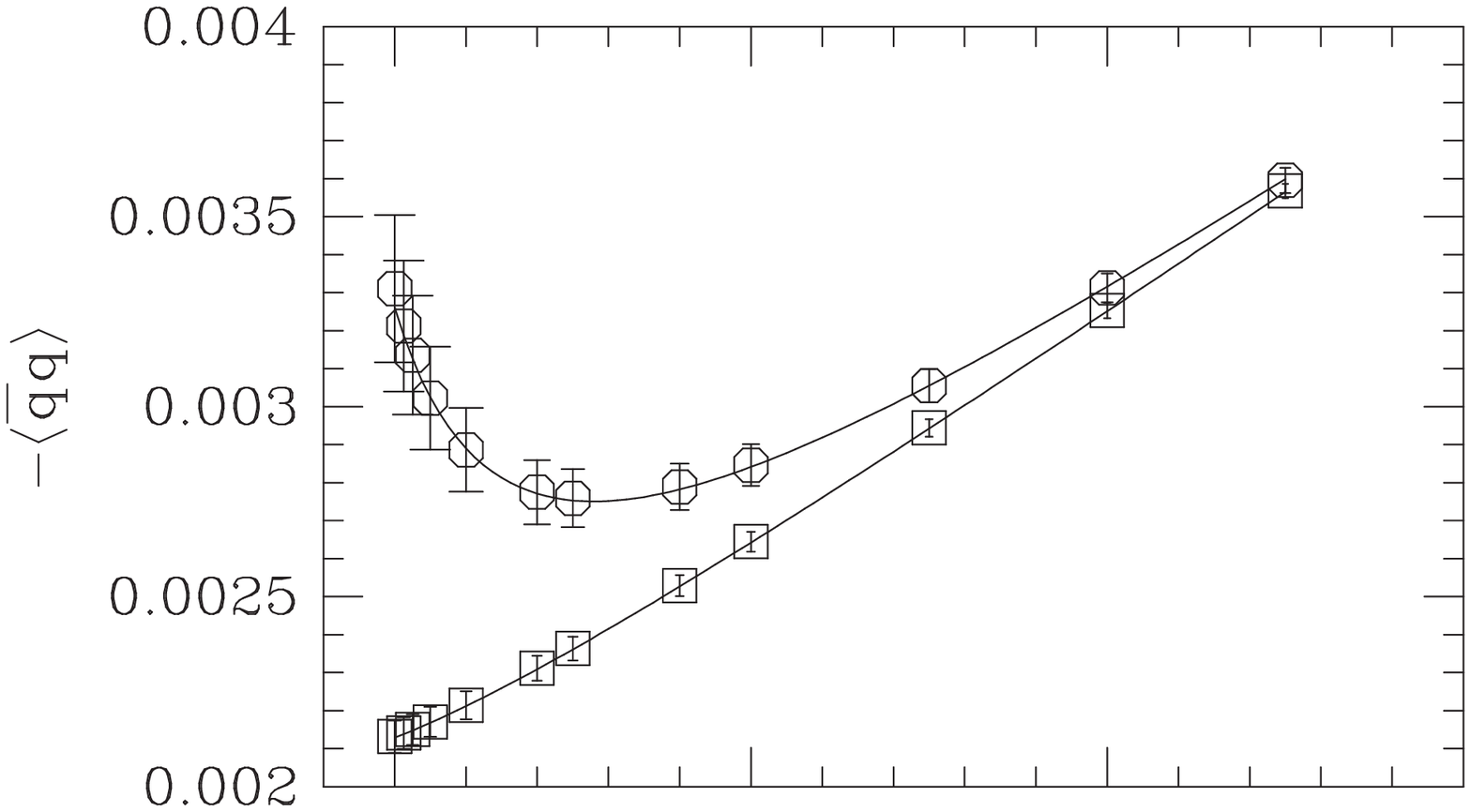}
\caption{
\label{fig:qbq_b57k165ns32}
Chiral condensate vs.\ mass for volumes $8^3\times 32$ (circles) and
$16^3\times 32$ (squares) at $\beta=5.7$, $L_s=32$.
}
\end{figure}

Fig.\ \ref{fig:qbq_b57k165ns32} shows $\qbq$
vs.\ the input quark mass $m_f$ on two volumes at
$\beta=5.7$ (plaquette gauge action).
As expected from Eq.\ \ref{eq:BC} the divergence is
much more severe for the smaller volume, approximately (1.6 fm)$^3$,
than for the larger volume, $\approx (3.2~{\rm fm})^3$.
The lines are fits to the form
\beq
  -\qbq = \frac{ a_{-1}}{ m_f + \dmqbq} + a_0 + a_1 m_f \, .
\eeq
$\qbq$ has a similar
divergence on a (1.6 fm)$^3$ box at $\beta=6.0$ \cite{Blum:2000kn}.

%%%%%
\section{PSEUDOSCALAR MESON}

\begin{figure}[t]
\epsfxsize=\hsize
\epsfbox{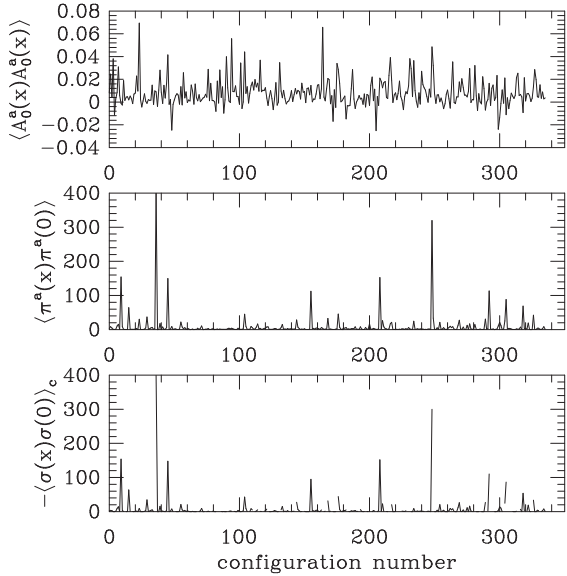}
\caption{
\label{fig:corr_832_b57k165ns48_m0.0}
Monte Carlo evolution of correlators on a $8^3\times32$ lattice
at $\beta=5.7$ with $L_s=48$ and $m_f=0.0$.
}
\end{figure}

Given that the chiral condensate in finite volume is 
sensitive to topological zero modes when using domain wall fermions,
we turn to correlators which couple to the pseudoscalar meson,
{\it viz} $\corraa$ and $\corrpp$.  Looking at the spectral
decomposition of these correlators \cite{Blum:2000kn}
leads one to expect the leading zero mode effects to be
$\corraa\propto 1/m$ and
$\corrpp \propto 1/m^2$.  Indeed time histories of these
correlators (Fig.\ \ref{fig:corr_832_b57k165ns48_m0.0})
show the latter correlator to be much more singular than
the former.  Also plotted in Fig.\ \ref{fig:corr_832_b57k165ns48_m0.0}
is the connected scalar--scalar correlator which
has the same leading zero mode effects as $\corrpp$, both
according to the spectral decomposition and as seen in the data.

\begin{figure}[t]
\epsfxsize=\hsize
\epsfbox{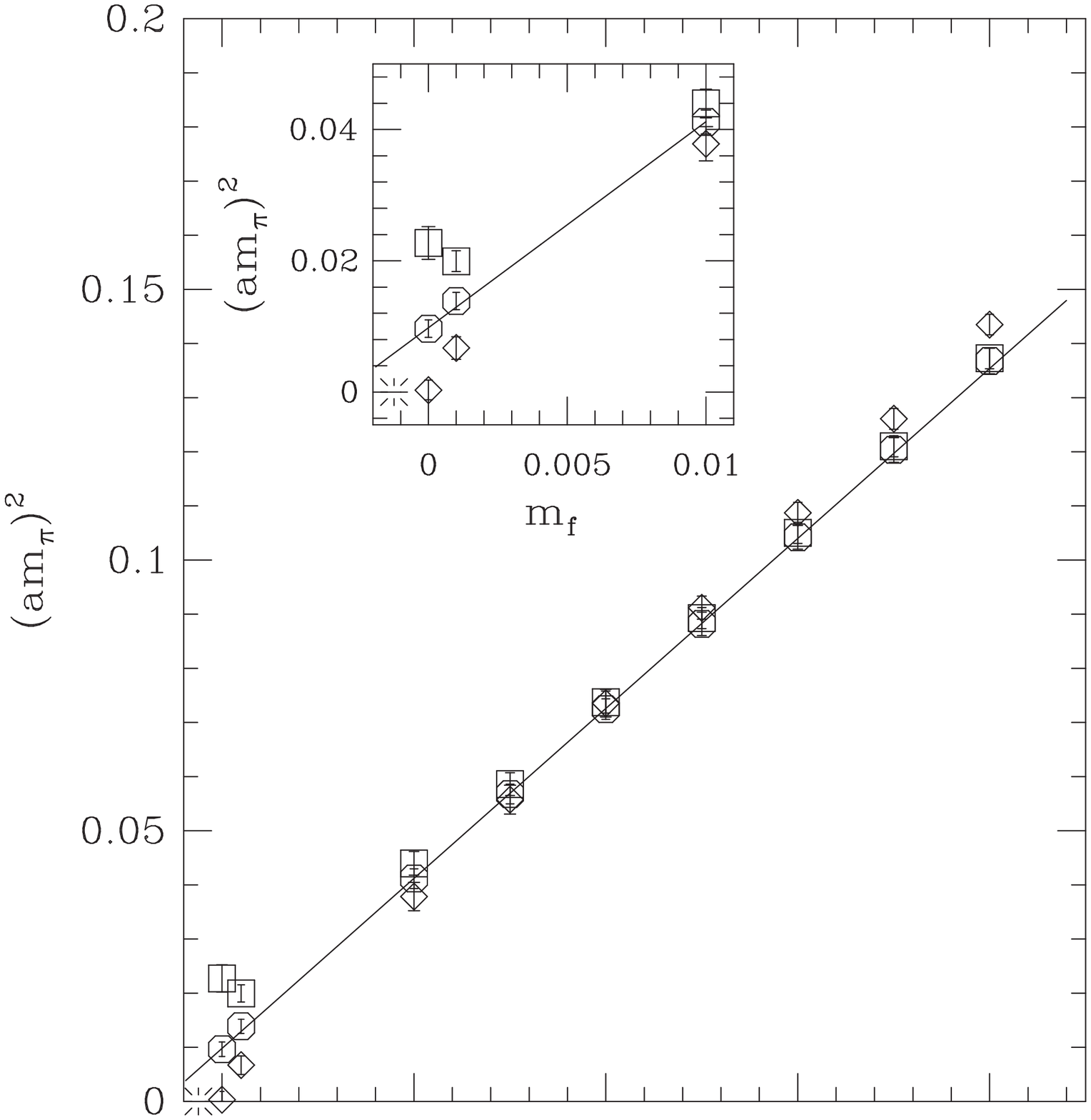}
\caption{
\label{fig:mpi2_1632_b60k18ns16}
Fits to pseudoscalar meson mass from $\corrpp$ ($\Box$),
$\corraa$ ($\circ$) and $\corrpp + \corrss$ ($\Diamond$) on a
$16^3\times 32$ lattice at $\beta=6.0$ with $L_s=16$.
The solid line is a linear fit to the circles with 
$0.01\le m_f \le 0.04$.
The asterisk marks $\mres$ and
the inset magnifies the data at small $m_f$.
}
\end{figure}

\begin{figure}[t]
\epsfxsize=\hsize
\epsfbox{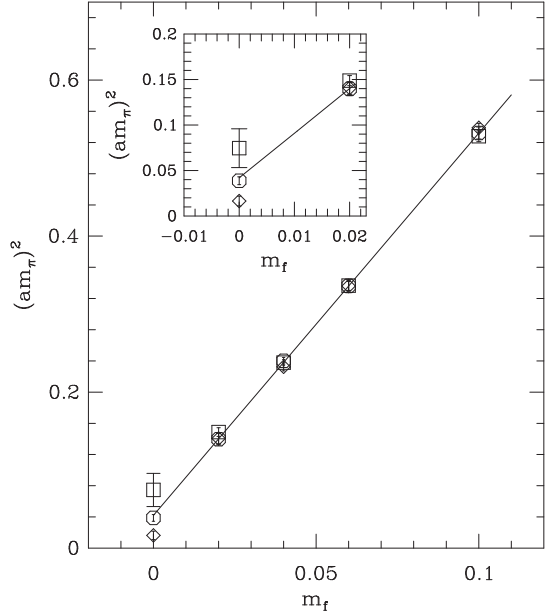}
\caption{
\label{fig:mpi2_832_b57k165ns48}
Fits to pseudoscalar meson mass from correlators denoted
as in Fig.\ \ref{fig:mpi2_1632_b60k18ns16} on a
$8^3\times 32$ lattice at $\beta=5.7$ with $L_s=48$.
The solid line is a linear fit to the circles
with $ 0.02 \le m_f \le 0.1$.  }
\end{figure}

The singular behavior of $\corrpp$ for small mass
leads to a contamination in the extraction of
the meson mass.  The mass computed from
this correlator is significantly larger at small
$m_f$ than the mass
computed from $\corraa$, both at $\beta=6.0$
(Fig.\ \ref{fig:mpi2_1632_b60k18ns16}) and at $\beta=5.7$ 
(Fig.\ \ref{fig:mpi2_832_b57k165ns48}).  An attempt can
be made to subtract the zero mode effects by extracting
a mass from $\corrpp + \corrss$.  At small $m_f$ the fact
that the mass from $\corraa$ is larger than from the
``subtracted'' correlator could be due to $1/m$ 
zero mode contributions in $\corraa$.  The subtracted
correlator receives increasing contribution from
the isovector scalar meson as $m_f$ increases, explaining
the deviation of the diamonds for larger values of $m_f$
in Fig.\ \ref{fig:mpi2_1632_b60k18ns16}.

\begin{figure}[t]
\epsfxsize=\hsize
\epsfbox{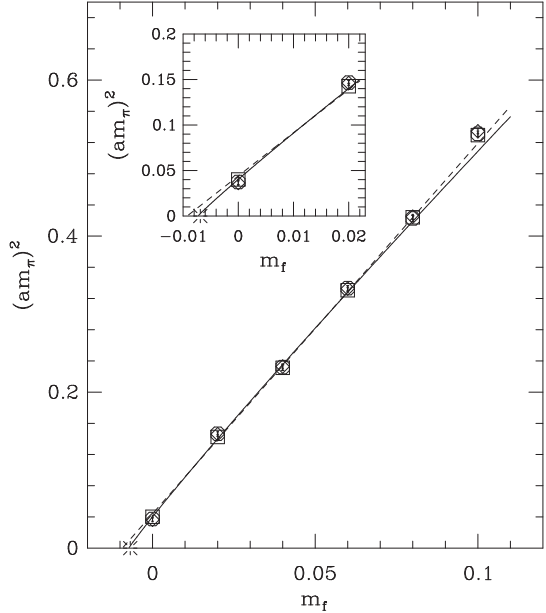}
\caption{
\label{fig:mpi2_1632_b57k165ns48_log}
Pseudoscalar meson mass squared vs.\ $m_f$ on a
$16^3\times 32$ lattice at $\beta=5.7$ with $L_s=48$.
Symbols used as in Fig.\ \ref{fig:mpi2_1632_b60k18ns16},
and the asterisk marks $m_{\rm res}$.
Solid line is a quenched chiral log 
fit to circles with $0.0 \le m_f \le 0.08$, and 
the dashed line is a linear fit to the same data.
}
\end{figure}

The zero mode effects disappear within 
the statistical resolution when the volume is increased.
Fig.\ \ref{fig:mpi2_1632_b57k165ns48_log} shows the pseudoscalar
meson mass extracted from all three propagators on a 
lattice with $(3.2~{\rm fm})^3$ spatial volume at $\beta =5.7$.
The asterisk in that figure denotes the residual mass, 
$\mres = 0.0072(9)$,
computed from the ratio of the mid-point pseudoscalar
density to the surface pseudoscalar density
\cite{Blum:2000kn,Wu:LAT00}.
Although it is not clear on the scale of the plot, the
linear extrapolation of $(am_\pi)^2$ to zero in $m_f$
misses $\mres$ by over $2\sigma$.  Quenched chiral perturbation
theory suggests that for vanishing quark mass $m$,
$m_\pi^2 \propto m^{1/(1+\delta)}$ \cite{Sharpe:1992ft}.
In order to simplify the
fit in the region of $m_f$ where the data points 
are, one uses
\beq
(am_\pi)^2 = a_0 ( m_f + a_1 ) ( 1 - \delta \ln(m_f + a_1 ) ) \, .
\eeq
The solid line in Fig.\ \ref{fig:mpi2_1632_b57k165ns48_log}
gives $a_1 = 0.0073(10)$, in good agreement with $\mres$, and
$\delta = 0.07(4)$.  Although the $\chi^2$/dof (with errors
estimated from jackknifing) decreases
from $4.3 \pm 2.6$ to $3.6 \pm 2.4$
between the linear and 
logarithmic fits, it is not compelling enough to 
conclude firmly that the latter is preferable.  On the other hand,
our theoretical prejudice is for a fit which extrapolates
to $(am_\pi)^2 = 0$ at $m_f = -\mres$.

The important conclusion from this study is that topological
zero modes can contaminate quantities used to compute
observables in small volume simulations.  This interesting
effect is a consequence of the index of the
domain wall Dirac operator and, and it 
vanishes for increasing volume.

%%%%%
\section{SCALING OF $m_N/m_\rho$}

\begin{figure}[t]
\vspace{7.5cm}
\includegraphics{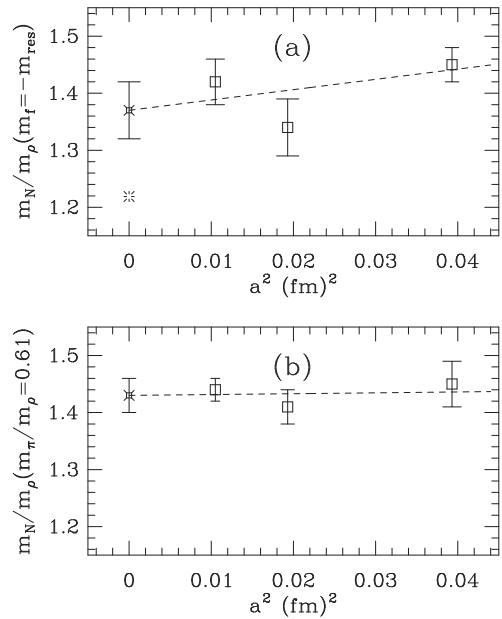}
\caption{
\label{fig:nuc_rho_vs_a2}
Nucleon--rho mass ratio on spatial volumes $\approx(1.6~{\rm fm})^3$.
Squares denote domain wall results at $\beta=$ 6.0, 5.85, and 5.7.
The chiral limit $m_f+\mres\to 0$ has been taken in (a), and
the asterisk marks the real world value.  (b) shows $m_N/m_\rho$
at $m_\pi/m_\rho=0.61$.  Dashed lines show fits to $c_0 + c_2 a^2$
and fancy squares the extrapolated $a=0$ value.
}
\end{figure}

Vector meson and nucleon masses have been computed
as functions of $m_f$ on lattices at $\beta = 5.7$, 5.85, 
and 6.0 on approximately equal spatial volumes.
Fig.\ \ref{fig:nuc_rho_vs_a2}a shows $m_N/m_\rho$ vs.\ $a^2$
where the nucleon and rho masses have been extrapolated
linearly to the chiral point $m_f = -\mres$.  Although
the scatter of the data points is only at the $1\sigma$
level, it is hard to draw a firm conclusion regarding
their scaling behavior.  The fit to $c_0 + c_2 a^2$
has a $\chi^2$/dof of 2.4.
%At $\beta = 5.7$, $m_N/m_\rho$ 
%decreases from 1.45(3) to 1.40(5) when the volume is 
%increased from $8^3$ to $16^3$, demonstrating that
%finite volume effects are a few percent at most.

Given that the data for the pseudoscalar
data hint at nonlinearities in the chiral extrapolations,
one might wonder what role these would play in 
Fig.\ \ref{fig:nuc_rho_vs_a2}a.  The smallest $m_f$ in
the $\beta = 5.7$, 5.85, and 6.0 simulations correspond to
$m_\pi/m_\rho = 0.38$, 0.57, and 0.46, respectively.  Therefore
the different simulations would have different sensitivities
to any deviation from linearity in the chiral limit.
One way to test scaling while avoiding chiral extrapolation
is to compare $m_N/m_\rho$ at a value of
$m_\pi/m_\rho$ where simulations have been done.
In Fig.\ \ref{fig:nuc_rho_vs_a2}b we plot $m_N/m_\rho$
interpolating the data to $m_\pi/m_\rho = 0.61$.
These data fit $c_0 + c_2 a^2$ with a $\chi^2$/dof of
0.88.

%%%%%
\section{STRANGE QUARK MASS}

The calculation of the strange quark mass using this data
was presented at DPF 2000 \cite{Wingate:DPF00QM}.
Using masses and matrix elements from \cite{Blum:2000kn}
the strange quark mass is computed defined through both the vector
and axial Ward-Takahashi identities (VWTI/AWTI).
The renormalization factors are computed in the RI/MOM scheme
\cite{Dawson:LAT00,RBC:Zfact}. We find at $\beta=6.0$
\beq
m_s^{\ol{\rm MS}}(2~{\rm GeV}) = \left\{
\begin{array}{cc}
110(2)(22)~{\rm MeV} & {\rm VWTI} \\
105(6)(21)~{\rm MeV} & {\rm AWTI} 
\end{array}
\right.
\eeq
using the $K$ mass to fix $m_f$ for the strange sector.
The first error is statistical, and the second is the
systematic error due to using perturbation theory at 2 GeV
to match to the $\ol{\rm MS}$ scheme.  At $\beta=5.85$
we quote $m_s^{\ol{\rm MS}}(2~{\rm GeV}) = 100(5)(20)$ MeV
from the AWTI.  Using the $K^*$ instead to set $m_f^{(s)}$
increases $m_s^{\ol{\rm MS}}(2~{\rm GeV})$ by
$\approx20\%$ at $\beta =6.0$ and by $\approx35\%$
at $\beta = 5.85$.

%%%%%
\section*{ACKNOWLEDGMENTS}

Simulations performed with the RBRC and Columbia QCDSP's
and the NERSC Cray T3E.

%%%%%%%%%%%%%%%%%%%%%%%%%%%

\end{document}